\documentstyle[preprint,aps]{revtex}
\def\beq{\begin{eqnarray}}
\def\eed{\end{eqnarray}}
\begin{document}
\draft

\title
{Electronic properties of the doped antiferromagnet on a kagom\'e lattice}

\author{Weiqiang Yu$^{1}$ and Shiping Feng$^{1,2}$}

\address{
$^{1*}$Department of Physics, Beijing Normal University, Beijing
 100875, China and \\
$^{2}$Institute of Theoretical Physics, Academia Sinica, Beijing 100080,
China \\
}


\maketitle

\begin{abstract}
Within the $t$-$J$ model, we study the electronic properties of the
doped antiferromagnet on the kagom\'e lattice based on the framework
of the self-consistent mean-field theory. At the half-filling, the
spin-liquid ground-state energy per site of the kagom\'e antiferromagnet
is $E_{g}/N_{s}J=-0.859$, which is in very good agreement with the
numerical estimates. Away from the half-filling, the electron
photoemission spectroscopy and density of states are discussed, and
the results indicate that there is a gap in the normal-state of the
system.
\end{abstract}
\pacs{71.20.Cf, 74.72.Hs, 79.60.Bm}


It has been clear in the last decades that detailed understanding of the
motion of the holes in a two-dimensional (2D) antiferromagnet is one of
the central issues in the theory of strongly correlated electron systems,
especially as it relates to the copper oxide superconductors \cite{1}.
This followed from the argument, made by many researchers \cite{2}, that
the essential physics of the copper oxide materials is contained in 2D
doped Mott insulators, obtained by chemically adding charge carriers to a
strongly correlated antiferromagnetic (AF) insulating state, therefore the
physical properties of these systems mainly depend on the extent of doping,
and the regimes have been classified into the underdoped, optimally doped,
and overdoped, respectively. The normal-state properties of the copper
oxide materials in the underdoped and optimally doped regimes exhibit a
number of anomalous properties in the sense that they do not fit in the
conventional Fermi-liquid theory \cite{3}. In particular, the unusual
normal-state gap behavior \cite{4}, and the stripe order of holes and
spins \cite{5} are found in the underdoped regime. It is also shown from
the experiments \cite{6} that the highest superconducting transition
temperature $T_{c}$ for the square lattice $CuO_2$ planes in the
tetragonal structure in the optimally doped regime is reduced by the 
structure distortions of the $CuO_2$ planes in the orthorhombic structure 
in the underdoped regime, and the stripe order of holes and spins is 
closely related with this distortion of the lattice \cite{5}. The spinon
frustration is also induced due to the distortion of the lattice of the
system. Motivated by the above experimental facts, we believe that the
doped kagom\'{e} lattice antiferromagnet (KLA) is an attractive candidate
for describing the distorted square $CuO_2$ planes since it has the same
coordinate number with the square lattice. On the other hand, the doped KLA,
which is the system with the geometric frustration as in the triangular 
lattice, is also of the theoretical interest in their own right, with many 
fascinating theoretical questions remaining unanswered. The undoped KLA
has been extensively studied \cite{7,8,9,10}, and the results show that its 
ground-state is the novel quantum disordered spin-liquid state without the 
AF long-range-order (AFLRO). It is also believed that the undoped KLA 
involving only nearest-neighbor coupling is the unique model to display the 
quantum spin liquid ground-state \cite{11}. Recognizing that these novel 
physics showed by the undoped KLA, a natural question is what is the 
electronic properties with doping on this system.

The situation for the doped square lattice antiferromagnet (SLA) is a bit 
more advanced. Some physical properties of the doped SLA have been studied 
intensively by many researchers within the $t$-$J$ model \cite{12}, where 
the strong electron correlation manifests itself by the electron single 
occupancy local constraint, and therefore the crucial requirement is to 
impose this local constraint in analytical calculations \cite{13}. Based on 
the charge-spin separation, it has been shown that a useful method to treat 
this local constraint is the fermion-spin theory \cite{14}. The fermion-spin 
theory has been used to discuss the physical properties of the doped SLA
\cite{15}, and the results are consistent with the experiments and numerical 
simulations. In this paper, we apply the fermion-spin approach \cite{14} to 
study the physical properties of the doped KLA. Our results show that at the 
half-filling the ground-state of the kagom\'{e} Heisenberg antiferromagnet 
is the quantum spin liquid state with energy per site $E_{g}/N_{s}J=-0.859$, 
which is in very good agreement with the numerical simulations 
\cite{7,8,9,10}. Away from half-filling, the electron photoemission 
spectroscopy and density of states are discussed, and the results indicate 
that there is a gap in the normal-state of the system.

The simple way to visualize the kagom\'{e} lattice is to regard the
triangular lattice as consisting of four sublattices and remove the spins
on one of the sublattices, therefore there are three inequivalent
sublattices $A$, $B$, and $C$ even without AFLRO, where the spins in $A$,
$B$, and $C$ sublattices pointing to the vertices of an equilateral
triangle are placed with no two nearest-neighbors pointing in the same
direction, then the kagom\'{e} lattice structure is much more complicated
than the square lattice. For convenience, the equilateral triangle $ABC$
is chosen as the cell as shown in Fig. 1, in this
case the kagom\'{e} lattice is reduced as the triangular structure, then
the position of the spin s in the cell $i$ is specified by the vector
$X_{i_{s}}=R_i+d_s$, with $s=A$, $B$, $C$, i.e., each cell contains
three spins, while the lattice vector $R_i$ and the reciprocal lattice
$K_j$ satisfy the relationship as $R_i\cdot K_j=2\pi \delta_{ij}$. With
the above definition, the 2D $t$-$J$ model on the kagom\'{e} lattice is
described by the Hamiltonian as,
\begin{eqnarray}
H=-t\sum\limits_{i_{s},\eta_{s},\sigma}C_{i_{s},\sigma}^{\dagger}
C_{i_{s}+\eta_{s},\sigma}+h.c.-\mu\sum\limits_{i_{s},\sigma}C_{i_{s},
\sigma}^{\dagger}C_{i_{s},\sigma}+J\sum\limits_{i_{s},\eta_{s}}
S_{i_{s}}\cdot S_{i_{s}+\eta_{s}},
\label{H}
\end{eqnarray}
where $C_{i_{s},\sigma}^{\dagger}(C_{i_{s},\sigma})$ are the electron
creation (annihilation) operators, $S_{i_{s}}=C_{i_{s}}^{\dagger}
\sigma C_{i_{s}}/2$ are spin operators with $\sigma=(\sigma_x,\sigma_y,
\sigma_z)$ as pauli operators, and $\mu$ is the chemical potential.
The sum is over all sites $X_{i_{s}}$, and for each $X_{i_{s}}$,
over the nearest-neighbor $\eta_{s}$. The Hamiltonian (\ref{H}) is
restricted to the sub-space where a given site cannot be occupied by 
more than one electron, i.e., $\sum_{\sigma}C_{i_{s},\sigma}^{\dagger}
C_{i_{s},\sigma}\leq 1$. In the fermion-spin representation \cite{14},
$C_{i\uparrow }=h_i^{\dagger}S_i^{-}$, $C_{i\downarrow }=h_i^{\dagger}
S_i^{+}$, where the spinless fermion operator $h_i$ keeps track of the 
charge (holon), while the pseudospin operator $S_i^{+}$ keeps track of 
the spin (spinon), the $t$-$J$ model (\ref{H}) is decoupled within the 
mean-field approximation (MFA) as \cite{15} $H=H_t+H_J-24Nt\chi\phi$ with
\begin{mathletters}
\begin{eqnarray}
H_t &=&2\chi t\sum\limits_{i_{s},\eta_{s}}h_{i_{s}+\eta_{s}}^{\dagger}
h_{i_{s}}-\mu\sum\limits_{i_{s}}h_{i_{s}}^{\dagger}h_{i_{s}}, \\
H_J &=&{1\over 2}J_{eff}\epsilon\sum\limits_{i_{s},\eta_{s}}(S_{i_{s}}^{+}
S_{i_{s}+\eta_{s}}^{-}+S_{i_{s}}^{-}S_{i_{s}+\eta_{s}}^{+})+J_{eff}\sum
\limits_{i_{s},\eta_{s}}S_{i_{s}}^zS_{i_{s}+\eta_{s}}^z,
\end{eqnarray}
\end{mathletters}
where N is the number of cells, $J_{eff}=J[(1-\delta)^2-\phi^2]$, and
$\epsilon =1+2t\phi/J_{eff}$. The nearest-neighbor spin bond-order
amplitude $\chi$ and holon particle-hole parameter $\phi$ are defined as
$\chi=\langle S_{i_{s}}^{+}S_{i_{s}+\eta_{s}}^{-}\rangle$, $\phi=\langle
h_{i_{s}}^{\dagger}h_{i_{s}+\eta_{s}}\rangle$, respectively. In the
present mean-field case, the holon moves in the background of spinons,
and the spinon's behavior is described by the anisotropic Heisenberg
model. Since there are three inequivalent sublattices $A$, $B$, and $C$
in the kagom\'{e} lattice system, then the one-particle spinon and holon
two-time Green's functions are defined as matrices,
\begin{eqnarray}
D_{\lambda}(k,t-t^{\prime}) &=&\left(\matrix{
D^{AA}_{\lambda}(k,t-t^{\prime}) &D^{AB}_{\lambda}(k,t-t^{\prime}) 
&D^{AC}_{\lambda}(k,t-t^{\prime})\cr
D^{BA}_{\lambda}(k,t-t^{\prime}) &D^{BB}_{\lambda}(k,t-t^{\prime}) 
&D^{BC}_{\lambda}(k,t-t^{\prime})\cr
D^{CA}_{\lambda}(k,t-t^{\prime}) &D^{CB}_{\lambda}(k,t-t^{\prime}) 
&D^{CC}_{\lambda}(k,t-t^{\prime})\cr
}\right),
\end{eqnarray}
where $\lambda=s$, $z$, $h$, $D^{\mu \nu}_s(i_{\mu}-j_{\nu},t-t^{\prime}
)=\langle\langle S_{i_{\mu}}^{+}(t);S_{j_{\nu}}^{-}(t^{\prime})\rangle
\rangle$ and $D_z^{\mu\nu}(i_{\mu}-j_{\nu},t-t^{\prime})=\langle\langle 
S_{i_{\mu}}^{z}(t);S_{j_{\nu}}^{z}(t^{\prime})\rangle\rangle$ with $\mu$, 
$\nu=A$, $B$, $C$, are the spinon's propagators, while 
$D^{\mu\nu}_h(i_{\mu}-j_{\nu},t-t^{\prime})=\langle\langle h_{i_{\mu}}(t);
h_{j_{\nu}}(t^{\prime})\rangle\rangle$ is the holon's propagator. At the 
half filling, the $t$-$J$ model is reduced as Heisenberg model. Since the 
absence of the simple three sublattice magnetic order on KLA has been 
convincingly demonstrated by many researchers \cite{11,16}, then in the 
following discussions, we will study the holon moves in the disordered 
spin liquid state, where there is no AFLRO, i.e., $\langle S_{i_{s}}^{z}
\rangle=0$. In this case, the basic equation for the Green's functions 
within $t$-$J$ model on the square lattice has been discussed in detail 
in Ref. \cite{15}. Following their discussions \cite{15}, the Green's 
functions (3) is obtained as,
\begin{eqnarray}
D_{\lambda}(k,\omega) &=&\sum_{j=1}^3\left(\matrix{
\Gamma^{\lambda}_{1j}(k,\omega)a^{\lambda}_{j1}(k) &\Gamma^{\lambda}_{1j}
(k,\omega)a^{\lambda}_{j2}(k) &\Gamma^{\lambda}_{1j}(k,\omega)
a^{\lambda}_{j3}(k)\cr
\Gamma^{\lambda}_{2j}(k,\omega)a^{\lambda}_{j1}(k) &\Gamma^{\lambda}_{2j}
(k,\omega)a^{\lambda}_{j2}(k) &\Gamma^{\lambda}_{2j}(k,\omega)
a^{\lambda}_{j3}(k)\cr
\Gamma^{\lambda}_{3j}(k,\omega)a^{\lambda}_{j1}(k) &\Gamma^{\lambda}_{3j}
(k,\omega)a^{\lambda}_{j2}(k) &\Gamma^{\lambda}_{3j}(k,\omega)
a^{\lambda}_{j3}(k)\cr
}\right)\times  \nonumber \\
&&{1\over [\omega^2-\omega_{\lambda 1}^2(k)][\omega^2-\omega_{\lambda 2}^2
(k)][\omega^2-\omega_{\lambda 3}^2(k)]}, 
\end{eqnarray}
where $\Gamma^{\xi}(k,\omega)$ is the adjugate matrix of
\begin{eqnarray}
\Delta_{\xi}(k,\omega) =\left(\matrix{
\omega^2-(Z_{\xi 1}+Z_{\xi 2}\gamma_{k1}) &Z_{\xi 3}\gamma_{k4}-Z_{\xi 2}
\gamma_{k5} &Z_{\xi 3}\gamma_{k6}-Z_{\xi 2}\gamma_{k7} \cr
Z_{\xi 3}\gamma_{k4}^{*}-Z_{\xi 2}\gamma_{k5}^{*} &\omega^2-(Z_{\xi 1}+
Z_{\xi 2}\gamma_{k2}) &Z_{\xi 3}\gamma_{k8}-Z_{\xi 2}\gamma_{k9}\cr
Z_{\xi 3}\gamma_{k6}^{*}-Z_{\xi 2}\gamma_{k7}^{*} &Z_{\xi 3}\gamma_{k8}^{*}
-Z_{\xi 2}\gamma_{k9}^{*} &\omega^2-(Z_{\xi 1}+Z_{\xi 2}\gamma _{k3})\cr
}\right),~~~~ 
\end{eqnarray}
with $\xi=s$, $z$, and $\Gamma^{h}(k,\omega)$ is the adjugate matrix of
\begin{eqnarray}
\Delta_h(k,\omega) =\left(\matrix{
\omega +\mu  &-4\chi t\gamma_{k4} &-4\chi t\gamma_{k6} \cr
-4\chi t\gamma_{k4}^{*} &\omega+\mu  &-4\chi t\gamma_{k8} \cr
-4\chi t\gamma_{k6}^{*} &-4\chi t\gamma_{k8}^{*} &\omega+\mu\cr
}\right) ,
\end{eqnarray}
while,
\begin{eqnarray}
a^{\xi}(k) =\left(\matrix{
-2Z_{\xi 4} &Z_{\xi 5}\gamma_{k4} &Z_{\xi 5}\gamma _{k6}\cr
Z_{\xi 5}\gamma_{k4}^{*} &-2Z_{\xi 4} &Z_{\xi 5}\gamma_{k8}\cr
Z_{\xi 5}\gamma_{k6}^{*} &Z_{\xi 5}\gamma_{k8}^{*} &-2Z_{\xi 4}\cr
}\right) , 
\end{eqnarray}
and $a^{h}=1$, where $Z_{s1}=8J_{eff}^2[\epsilon^2(\alpha C+1/2)+(\alpha 
C_z+1/2)]$, $Z_{s2}=4J_{eff}^2(\epsilon^2\alpha \chi_z+\epsilon\alpha\chi)$, 
$Z_{s3}=4J_{eff}^2[\epsilon(\alpha C+\alpha C_z+1)+3(\epsilon^2\alpha\chi
+\epsilon\alpha\chi_z)-(\epsilon^2\alpha\chi_z+\epsilon\alpha\chi)]$,
$Z_{s4}=4J_{eff}(\epsilon\chi+\chi_z)$, $Z_{s5}=4J_{eff}(\epsilon
\chi_z+\chi)$, $Z_{z1}=16J_{eff}^2\epsilon^2(\alpha C+1/2)$,
$Z_{z2}=8J_{eff}^2\epsilon\alpha\chi$, $Z_{z3}=Z_{21}/2+2Z_{22}$,
$Z_{z4}=4J_{eff}\chi$, $Z_{z5}=Z_{z4}$, $\gamma_{k1}={\rm cos}k_y+{\rm cos}
(k_y-k_x)$, $\gamma_{k2}={\rm cos}k_x+{\rm cos}k_y$, $\gamma_{k3}={\rm cos}
k_x+{\rm cos}(k_y-k_x)$, $\gamma_{k4}=(1+e^{ik_y})/2$, $\gamma_{k5}=(e^{ik_x}
+e^{i(k_y-k_x)})/2$, $\gamma_{k6}=(1+e^{i(k_y-k_x)})/2$,  $\gamma_{k7}=
(e^{-ik_x}+e^{ik_y})/2$,  $\gamma_{k8}=(1+e^{-ik_x})/2$,  $\gamma _{k9}
=(e^{-ik_y}+e^{i(k_y-k_x)})/2$, and the order parameters $C=\sum\limits
_{\eta_{s}\neq\eta_{s^{\prime}}}\langle S_{i_{s}+\eta_{s}}^{+}S_{i_{s}+
\eta_{s^{\prime}}}^{-}\rangle$, $\chi_z=2\langle S_{i_{s}}^zS_{i_{s}+
\eta_{s}}^z\rangle$, $C_z=2\sum\limits_{\eta_{s}\neq\eta_{s^{\prime}}}
\langle S_{i_{s}+\eta_{s}}^zS_{i_{s}+\eta_{s^{\prime}}}^z\rangle$, while
$\omega_{\lambda j}(k)$ is the solution of the determinant 
$|\Delta_{\lambda}(k,\omega_j)|=0$. In order not to violate the
sum rule $\langle S_{i_{s}}^{+}S_{i_{s}}^{-}\rangle =1/2$ in the case
$\langle S_{i_{s}}^z\rangle =0$, the important decoupling parameter
$\alpha$ has been introduced in the above calculation, which can be
regarded as the vertex corrections \cite{15,17}. With the help of the
spectral representation of the correlation functions, the decoupling 
parameter $\alpha$ and order parameters $\chi$, $\chi_z$, $C$, $C_z$, 
$\phi$, and chemical potential $\mu$ can be determined by the seven 
self-consistent equations \cite{15}.

At the half-filling, the $t$-$J$ model is reduced as the Heisenberg model, 
where $\epsilon=1$, $\chi_z=\chi$, $C_z=C$ in the rotational symmetrical 
case, and therefore $D_z(k,\omega)=D_s(k,\omega)/2$. In this case, the 
disordered spin liquid ground-state energy per site is $E_g/N_{s}J=-0.859$. 
For comparison, some results of the ground-state energy of KLA obtained 
from the numerical simulations and the present theoretical result are 
listed in Table I, where we find that the present spin liquid energy seems 
to be closer to the extrapolated finite lattice results of Leung and Elser 
\cite{8}, only is more than $3\%$ higher than the best numerical estimates 
by Zeng and Elser \cite{7}, and is almost identical with that of the 
variational calculation by Sindzinger, Lecheminant and Lhuillier \cite{10}. 
Therefore the accuracy of the present mean-field results at half-filling is 
confirmed. As a by-product, the Heisenberg antiferromagnet on the triangular 
lattice has been discussed, and the spin liquid energy per site is 
$E_g/NJ=-0.966$, which is essentially identical to results obtained by 
Kalmeyer and Laughlin \cite{18} and Lee and Feng \cite{19} based on the 
resonating-valence-bond state.

Away from half-filling, we find that the phase separation, present in the 
doped SLA \cite{20}, is absent here. For discussing the physical properties 
of the electronic state of the doped KLA, we need to calculate the electron 
Green's function $G^{\mu\nu}(i_{\mu}-j_{\nu},t-t^{\prime})=\langle\langle 
C_{i_{\mu}\sigma}^{\dagger}(t);C_{j_{\nu}\sigma}(t^{\prime})\rangle\rangle$, 
which is a convolution of the spinon Green's function $D_s(k,t-t^{\prime})$ 
and holon Green's function $D_h(k,t-t^{\prime})$ in the framework of the
fermion-spin theory, and can be obtained at the mean-field level as,
\begin{eqnarray}
G^{\nu A}(k,\omega) ={1\over N}\sum\limits_{\vec{p}}\int
\nolimits_{-\infty }^\infty {d\omega^{\prime}\over 2\pi}\int
\nolimits_{-\infty }^\infty {d\omega^{\prime\prime}\over 2\pi}{n_F
(\omega^{\prime})+n_B(\omega^{\prime\prime})\over \omega+\omega^{\prime}
-\omega^{\prime\prime}}
\left( 
\begin{array}{c}
A_h^{A\nu}(p-k,\omega^{\prime})A_s^{\nu A}(p,\omega^{\prime\prime}) \\
A_h^{A\nu}(p-k,\omega^{\prime})A_s^{\nu A}(p,\omega^{\prime\prime}) \\
A_h^{A\nu}(p-k,\omega^{\prime})A_s^{\nu A}(p,\omega^{\prime\prime})
\label{electron}
\end{array}
\right), 
\end{eqnarray}
where $A_h^{\mu\nu}(k,\omega)=-2{\rm Im}D_h^{\mu\nu}(k,\omega)$ and
$A_s^{\mu\nu}(k,\omega)=-2{\rm Im}D_s^{\mu\nu}(k,\omega)$ are the holon
and spinon spectral functions, respectively, $n_B(\omega)$ is the boson
distribution function, while $n_F(\omega)$ is the fermion distribution
function. From the Green's function (\ref{electron}), we obtain the
electron spectral function as $A(k,\omega) =-2{\rm Im}G^{AA}(k,\omega)-
2{\rm Im}G^{BA}(k,\omega)-{\rm Im}G^{CA}(k,\omega)$.

Although the structural distortions of the $CuO_2$ planes in the copper
oxide materials in the underdoped regime, measured most conveniently by
the copper-oxygen bond angles \cite{6}, are very complicated and materials
dependent, the common feature is that the spin frustration has been 
introduced. This spin frustration is related to the variation of the 
electronic density of states at the Fermi energy \cite{6}. Among the doped
antiferromagnets the most helpful for discussing the effects of the spin
frustration on the electronic properties due to the distortions
of the square $CuO_2$ planes may be the doped KLA since, as mentioned
above, it has the same coordinate number with the square lattice. We have
performed the numerical calculation for the electron spectral function
$A(k,\omega)$ of the doped KLA, and the results at the doping $\delta =0.12$ 
for the parameter $t/J=2.5$ at the zero temperature is plotted in Fig. 2. 
In comparison with the results of the spectral function on the square 
lattice \cite{15}, it is shown that the structure of the spectroscopy on 
the kagom\'{e} lattice is much complicated than the square lattice. Moving 
in the momentum space in the direction of O point, the photoemission 
spectroscopy weight increases, while the inverse photoemission spectroscopy 
weight decreases. On the contrary, increasing momentum towards the $M$ 
point, the situation is inversed. For the further understanding of the 
electronic state properties, we investigate the electron density of states 
$\rho(\omega)$, which is closely related with the spectral function, and is 
given as $\rho(\omega)={1\over N}\sum\limits_{k}A(k,\omega)$. The results
of $\rho(\omega)$ with (a) the doping $\delta =0.12$ and (b) the doping
$\delta =0.06$ for the parameter $t/J=2.5$ at the zero temperature are
plotted in Fig. 3, where the existence of the gap in the electron density
of states is an important feature. It is also shown that the density of
states is shifted towards smaller energies with increasing doped holes,
and the total weight is reduced since the integral of the density of
states up to the Fermi energy has to be equal to the number of electrons,
therefore there is a tendency that the gap narrows with increasing dopings
since some states appear in the gap upon dopings. These results are
consistent with the experiments of the copper oxide materials in the
underdoped regime \cite{4}. In comparison with the results of the density
of states on the square lattice \cite{15}, we find that the electron
density of states near the Fermi energy on the kagom\'{e} lattice is
lower than these on the square lattice. Irrespective of the coupling
mechanism responsible for the superconductivity in the copper oxide
materials, the transition temperature $T_c$ is usually related to the
electronic density of states at the Fermi energy. Our results may
interpret that the highest superconducting transition temperature $T_{c}$
for the square lattice $CuO_2$ planes in the tetragonal structure in the
optimal doping is reduced by the structure distortions of the $CuO_2$
planes in the orthorhombic structure in the underdoped regime \cite{6},
i.e., this observed decrease of $T_c$ in the underdoped regime \cite{6}
is due to the split singularity of the electronic density of states with
the gap and lower electronic density of states near the Fermi energy.
Moreover, the present results also indicate that the existence of the
normal state gap and decrease of the electronic density of state near
Fermi energy is possibly induced by the distortions of the lattice
(then the spinon frustration).

In summary, we have discussed the electronic properties of the doped KLA
within the $t$-$J$ model. In the self-consistent mean-field level, we show 
that the ground-state of the undoped KLA is the spin liquid state with the 
energy per site $E_g/N_{s}J=-0.859$, which is consistent with the numerical 
simulations \cite{7,8,9,10}. It is also indicated that there is the 
normal-state gap in the electronic density of states for the doped KLA.

\acknowledgments
This work was supported by the National Natural Science Foundation under
the Grant No. 19774014 and the State Education Department of China through 
the Foundation of Doctoral Training. The partial support from the Earmarked
Grant for Research from the Research Grants Council (RGC) of the Hong Kong,
China is also acknowledged.

\begin{figure}
\caption{The spin configuration of the sublattices $A$, $B$, and $C$ on 
the kagom\'e lattice.}
\end{figure}

\begin{figure}
\caption{Electron spectral function $A(k,\omega)$ of the $t$-$J$ model 
on the kagom\'e lattice within the self-consistent mean-field theory in 
the doping $\delta =0.12$ for the parameter $t/J=2.5$ at the temperature
$T=0$. Note that there is a gap in the electron spectroscopy.}
\end{figure}

\begin{figure}
\caption{Electron spectral density of the $t$-$J$ model  on the kagom\'e
lattice in the doping (a) $\delta =0.12$ and (b) $\delta =0.06$ for
the parameter $t/J=2.5$.}
\end{figure}

\begin{table}
\caption{A comparison of the ground-state energy per site for the 
antiferromagnetic Heisenberg model on the two-dimensional kagom\'e
lattice.}
\begin{tabular}{|c|l|l|}
Authors           &   $E_g/N_sJ$  &   Method \\
\hline
Zeng and Elser \cite{7}   & $-0.882$  &  Finite lattice \\
\hline 
Yang, Warman and Girvin \cite{9}   & $-0.788$  &  Variational Monte Carlo \\
\hline 
Leung and Elser \cite{8}   & $-0.877$  &  Finite lattice \\
\hline 
Sindzingre, Lecheminant and Lhuillier \cite{10}   & $-0.84$  &  Variational
Monte Carlo \\
\hline 
The present work   & $-0.859$  &  Green's function method \\
\end{tabular}
\end{table}

\end{document}